\documentclass[11pt]{article}

\usepackage{amssymb} \usepackage{amsfonts} \usepackage{amsmath} \usepackage{mathrsfs} \usepackage[dvips]{graphicx} \usepackage{color}

\begin{document}

\title{Strong coupling limits and quantum isomorphisms of the gauged Thirring model} 
\author{R. Bufalo$^1$\thanks{%
rbufalo@ift.unesp.br}, R. Casana$^2$\thanks{%
casana@ufma.br}~ and B.M. Pimentel$^1$\thanks{%
pimentel@ift.unesp.br}~ \\
\textit{{$^{1}${\small Instituto de F\'{\i}sica Te\'orica (IFT/UNESP), UNESP
- S\~ao Paulo State University}}} \\
\textit{\small Rua Dr. Bento Teobaldo Ferraz 271, Bloco II Barra
Funda, CEP 01140-070,S\~ao Paulo, SP, Brazil}\\
\textit{{\small $^2$ Departamento de F\'{\i}sica, Universidade
Federal do
Maranh\~{a}o (UFMA),}} \\
\textit{{\small Campus Universit\'{a}rio do Bacanga, CEP 65085-580, S\~{a}o
Lu\'{\i}s - MA, Brasil.}}}
\date{}
\maketitle

\begin{abstract} We have studied the quantum equivalence in the respective strong coupling limits of the bidimensional gauged Thirring model with both Schwinger
and Thirring models. It is achieved following a nonperturbative quantization of the gauged Thirring model into the path-integral approach. First, we have
established the constraint structure via the Dirac's formalism for constrained systems and defined the correct vacuum--vacuum transition amplitude by using the
Faddeev-Senjanovic method. Next, we have computed exactly the relevant Green's functions and shown the Ward-Takahashi identities. Afterwards, we have established
the quantum isomorphisms between gauged Thirring model and both Schwinger and Thirring models by analyzing the respective Green's functions in the strong
coupling limits, respectively. A special attention is necessary to establish the quantum isomorphism between the gauged Thirring model and the Thirring model.
\end{abstract}

\maketitle

\section{Introduction}

In general, in (3+1)-dimensions, it is well-known that the quantization procedures do not solve exactly any interacting quantum field theory especially a gauge
field theory. However, in low-dimensional space-times \cite{3} many interesting non-perturbative features of quantum field theory can be analyzed, for example,
in (1+1)-dimensions some models as quantum electrodynamics and fermionic quartic interactions become exactly solvable. Also, in the last years there has been a
growing interest in exactly solvable low-dimensional quantum field models due to their applicability in some problems of Condensate Matter Physics
\cite{fradkin1,fradkin2,fradkin3}.

In the context of quantum field theory, the dimensional models have been widely explored to test various relevant phenomena in more realistic models, such as
dynamical mass generation, asymptotic freedom and confinement. The study and search of solvable models in quantum field theory was begun with the proposals of W.
Thirring and J. Schwinger. For example, Schwinger \cite {5} has shown two important features, that it is not necessary a massless gauge field to preserve the
local gauge symmetry and that the fermionic field is confined, in total analogy with the quark confinement phenomenon happening in quantum chromodynamics
(QCD$_{4}$).

The Thirring model \cite{6} (TM) describes a self-interaction of massless Dirac's fermion fields in $(1+1)-$ dimensions and some exact solutions of model were
carried out by B. Klaiber \cite{7} and by N. Nakanishi \cite{8a,8b}. It is well-known that the Thirring model does not present a local gauge invariance, however,
at quantum level, it has been shown that TM would have a sector with a explicit local gauge symmetry \cite{9}. The possibility of a gauge invariant TM was
explored firstly by Itoh, et al., \cite{10} such reformulation was performed by using the Hidden Local Symmetry technique which gives a gauge field character,
with coupling constant $e$, to the auxiliary vectorial field $A_{\mu }$ which linearizes the fermionic self-interaction of the original TM. The Hidden Local
Symmetry technique is also known as the St\"{u}ckelberg method\footnote{For a review of the St\"{u}ckelberg's formalism see \cite{11}.}.

Another proposal for a gauged Thirring model (GTM) was suggested by K. Kondo in \cite{12a,12b,12c} where it has been discussed and studied several features of
the GTM, including the issue of fermionic mass generation. However, one of the most interesting analysis was the study at classical level of its behavior in the
strong  coupling regime. Such analysis performed in the unitary gauge allows to reproduce both the Schwinger and Thirring models.

Also, many aspects of  {gauged Thirring model} such as classical theory, Thermodynamics, generating functional, bosonization, chiral condensate, relation to
Schwinger model, in curved space-time, etc., have been extensively investigated in Refs. \cite{wipf1} and \cite{wipf2}.

Although through years many properties of GTM have been studied, the quantum analysis of the strong coupling limits of the model has been not performed.
Therefore, the aim of the present work is to analyze at quantum level the  {strong coupling limits} by using functional techniques. For this goal we will derive
the corresponding Green's functions and the respective Ward-Takahashi (WT) identities \cite{13}. As results of our analysis, we find  {the quantum GTM, in} its
respective strong coupling limits, reproduces as much the quantum SM as the quantum TM. Nevertheless, it is necessary a more careful analysis to obtain the TM
sector, because in the quantum TM does not exist a dynamical gauge field but the fermionic current $ j^{\mu }=\bar{\psi}\gamma ^{\mu }\psi $ does it. Therefore,
the correlation functions involving only the fermionic current $j^{\mu }$ and the fermionic field in GTM have its quantum correspondence with TM in the
respective strong coupling regimen.

This paper is organized as follow: In Sec. 2, we study the constraint structure of GTM which includes the constraint classification and the imposition of
noncovariant gauge conditions. In Sec. 3, we construct the vacuum-vacuum transition amplitude for GTM in the covariant $R_{\xi }-$gauge and compute the relevant
Green's functions and the Ward-Takahashi identities. In Sec. 4, we will go beyond Kondo's suggestion and we establish the strong coupling limits of the gauge
Thirring model at quantum level. In the last section we present our final remarks and perspectives.

\section{The gauged Thirring model: Hamiltonian structure}

The gauged Thirring model is described by the following Lagrangian density \begin{equation} \mathcal{L}=\frac{i}{2}\bar{\psi}\overleftrightarrow{\partial
\!\!\!/}\psi -m \bar{\psi}\psi +\bar{\psi}A\!\!\!/\psi +\frac{1}{2g}\left( A_{\mu }-\partial _{\mu }\theta \right) ^{2}-\frac{1}{4e^{2}}F_{\mu \nu }F^{\mu \nu
}, \label{eq 1.1} \end{equation} where the field-strength tensor $F_{\mu \nu }= \partial _{\mu } A_{\nu }- \partial _{\nu }A_{\mu }$, and $e>0$. The $\theta $
field was introduced following the St\"{u}ckelberg procedure. At classical level, the Lagrangian density is invariant under the following local gauge $U\left(
1\right) $ symmetry: \begin{eqnarray} &\psi ^{\prime }\left( x\right) =e^{i\lambda \left( x\right) }\psi \left( x\right) \;,\;\, \bar{\psi}^{\prime }\left(
x\right) =\bar{\psi}\left( x\right) e^{-i\lambda \left( x\right) },&  \notag \\[-0.2cm] & &  \label{eq 1.2} \\[-0.2cm] &A_{\mu }^{\prime }\left( x\right) =A_{\mu
}\left( x\right) +\partial _{\mu }\lambda \left( x\right) \;,\;\, \theta ^{\prime }\left( x\right) =\theta \left( x\right) +\lambda \left( x\right) . &  \notag
\end{eqnarray}

The Euler-Lagrange equations are given by \begin{eqnarray} \left( i\partial \!\!\!/+A\!\!\!/\left( x\right) -m\right) \psi \left( x\right) &=&0,  \label{eq 1.3}
\\[0.2cm] \frac{1}{e^{2}}\partial _{\mu }F^{\mu \nu }+\bar{\psi}\left( x\right) \gamma ^{\nu }\psi \left( x\right) +\frac{1}{g}\left( A^{\nu }\left( x\right)
-\partial ^{\nu }\theta \left( x\right) \right) &=&0,  \label{eq 1.5} \\ [0.2cm] \partial _{\mu }A^{\mu }\left( x\right) -\square \theta \left( x\right) &=&0.
\label{eq 1.6} \end{eqnarray}

Starting from the field equations and following Kondo's suggestion, we see the
 {strong coupling limits} in the unitary gauge ( $\theta =0$):

\begin{itemize} \item In the limit $g\rightarrow \infty $, we recover the field equations of massive  Schwinger model,\emph{\ i. e.}, $\mathcal{L} \rightarrow
\mathcal{L} _{SM}$.

\item In the limit $e\rightarrow \infty $, we recover the field equations of massive  Thirring model, \emph{i. e.}, $\mathcal{L}\rightarrow\mathcal{L}_{TM}$.
     \end{itemize}

In order to accomplish the Hamiltonian analysis of this model, we begin to define the canonical conjugate momenta to the field variables \begin{eqnarray}
\bar{p}_{a} &= &\frac{\partial \mathcal{L}}{\partial \left( \partial _{0}\psi _{a}\right) }=-\frac{i}{2}\bar{\psi}_{b}\left( \gamma ^{0}\right) _{ba},  \label{eq
1.7} \\[0.2cm] p_{a} &= &\frac{\partial \mathcal{L}}{\partial \left( \partial _{0}\bar{\psi} _{a}\right) }=-\frac{i}{2}\left( \gamma ^{0}\right) _{ab}\psi _{b},
\label{eq 1.8} \\[0.2cm] \pi ^{\mu } &= &\frac{\partial \mathcal{L}}{\partial \left( \partial _{0}A_{\mu }\right) }=\frac{1}{e^{2}}F^{\mu 0},  \label{eq 1.9}
\\[0.2cm] \pi _{\theta } &= &\frac{\partial \mathcal{L}}{\partial \left( \partial _{0} \theta \right) }=-\frac{1}{g}A_{0}+\frac{1}{g}\partial _{0}\theta .
\label{eq 1.10} \end{eqnarray}

The constraint analysis procedure \cite{14a,14b,14c} tells us that the momentum expressions given by Eqs. (\ref{eq 1.7}), (\ref{eq 1.8}) and (\ref{eq 1.9}) yield
three primary constraints \begin{eqnarray} \phi _{a} &= &p_{a}+\frac{i}{2}\left( \gamma ^{0}\right) _{ab}\psi_{b}\approx 0,  \label{eq 1.11} \\[0.2cm]
\bar{\phi}_{a} &= &\bar{p}_{a}+\frac{i}{2}\bar{\psi}_{b}\left( \gamma ^{0}\right) _{ba}\approx 0,  \label{eq 1.12} \\[0.2cm] \varphi _{1} &= &\pi ^{0}\approx 0.
\label{eq 1.13} \end{eqnarray} On the other hand, from Eqs.(\ref{eq 1.9}) and (\ref{eq 1.10}), we obtain two dynamical relations \begin{eqnarray} \partial
_{0}A_{1} &=&e^{2}\pi ^{1}+\partial _{1}A_{0},  \label{eq 1.14} \\ [0.2cm] \partial _{0}\theta &=&g\pi _{\theta }+A_{0}.  \label{eq 1.15} \end{eqnarray}

Along the paper we use the left derivative concept \cite{15} by dealing with grassmannian variables (fermionic field components).

Now we can write out the canonical Hamiltonian density as \begin{eqnarray} \mathcal{H_{C}} &=&-\frac{i}{2}\bar{\psi}\gamma ^{1}\overleftrightarrow{ \partial
}_{1}\psi +m\bar{\psi}\psi -\bar{\psi}A\!\!\!/\psi +\frac{g}{2} \left( \pi _{\theta }\right) ^{2}+\frac{e^{2}}{2}\left( \pi ^{1}\right) ^{2}+\pi ^{1}\left(
\partial _{1}A_{0}\right) +  \notag \\[-0.2cm] &&  \label{eq 1.16} \\[-0.2cm] &&+A_{0}\pi _{\theta }+\frac{1}{2g}\left( A_{1}\right) ^{2}+\frac{1}{2g} \left(
\partial _{1}\theta \right) ^{2}-\frac{1}{g}A_{1}\left( \partial _{1}\theta \right) .  \notag \end{eqnarray}

Following the usual Dirac procedure, we introduce the primary Hamiltonian ($ H_{P}$) by adding to the canonical Hamiltonian all the primary constraints,
\begin{equation} H_{P}=H_{C}+\int dz^{1}\left( \bar{\phi}_{a}\lambda _{a}+\bar{\lambda} _{a}\phi _{a}+v_{1}\varphi _{1}\right) ,  \notag \end{equation} where
the $\lambda ,\bar{\lambda}$ are fermionic Lagrange multipliers and $ v_{1}$ is a bosonic Lagrange multipliers, all them to be determined.

The fundamental Berezin brackets (BB) of theory are \begin{eqnarray} \left\{ \psi _{a}\left( x\right) ,\bar{p}_{b}\left( y\right) \right\} _{B} &=&-\delta
_{ab}\delta\!\left( x^{1}-y^{1}\right) ,  \label{eq 1.17} \\[0.2cm] \left\{ \bar{\psi}_{a}\left( x\right) ,p_{b}\left( y\right) \right\} _{B} &=&-\delta
_{ab}\delta\! \left( x^{1}-y^{1}\right) ,  \label{eq 1.18} \\[0.2cm] \left\{ A^{\mu }\left( x\right) ,\pi _{\nu }\left( y\right) \right\} _{B}
&=&\delta^\mu{}_\nu\delta \!\left( x^{1}-y^{1}\right) , \label{eq 1.19} \\[0.2cm] \left\{ \theta \left( x\right) ,\pi _{\theta }\left( y\right) \right\} _{B}
&=&\delta \!\left( x^{1}-y^{1}\right) .  \label{eq1.20} \end{eqnarray} It is easy to see that the only non-null BB between the primary constraints is
\begin{equation} \left\{ \phi _{a}\left( x\right) ,\bar{\phi}_{b}\left( y\right) \right\} _{B}=-i\left( \gamma ^{0}\right) _{ab}\delta \left( x^{1}-y^{1}\right)
. \label{eq 164} \end{equation}

The full constraint analysis \cite{14a,14b,14c} yields a set $\left\{ \varphi _{1}, \varphi _{2}\right\} $ of first--class constraints, where the constraint $
\varphi _{2}$ is given by \begin{equation} \varphi _{2}=\partial _{1}\pi ^{1}-\pi _{\theta }+i\left( \bar{p}\psi +\bar{ \psi}p\right) \approx 0,  \label{eq
1.21} \end{equation} it is Gauss's law: $\partial _{1}\pi ^{1}+\bar{\psi}\gamma^{0}\psi -\pi _{\theta }=0$. Also, we obtain from such analysis the second--class
set given by $\left\{ \phi _{\alpha },\bar{\phi}_{\alpha }\right\} $.

As the theory presents first--class constraints, \emph{i. e.} gauge symmetry generators, the Hamiltonian remains undetermined. Therefore, to define an unique
Hamiltonian it is necessary to introduce two additional conditions which are named as gauge fixing conditions. In that way, we choose the gauge condition known
as the radiation gauge \begin{eqnarray} \chi _{1} &=&A_{0}\approx 0,  \label{eq 1.24} \\[0.2cm] \chi _{2} &=&\partial _{1}A_{1}\approx 0.  \label{eq 1.25}
\end{eqnarray}

Once the Hamiltonian structure has been fully determined via the Dirac's method we can quantize correctly the model following the Faddeev-Senjanovic procedure
\cite{2a,2b,2c} to implement the functional quantization of the gauged Thirring model, as will be done in next section.

\section{The vacuum--vacuum transition amplitude}

In this section we will proceed with the path--integral quantization of gauged Thirring model. As we are dealing with a constrained system, we need to proceed
carefully with the functional quantization process. Thus, after determining and classifying the complete set of constraints including the imposition of the gauge
fixing conditions, we are in conditions of applying Faddeev-Senjanovic's method to accomplish the transition amplitude in a correct way. Therefore, the
vacuum--vacuum transition amplitude in its Hamiltonian form is given by \begin{eqnarray} Z=\int \mathcal{D}\mu \left[ {\psi ,\bar{\psi},A_{\mu },\theta }\right]
\exp  \Big( i\int d^{2}x\Big[ \pi ^{\mu }\left( \partial _{0}A_{\mu }\right) +\pi _{\theta }\left( \partial _{0}\theta \right) +\left( \partial_{0} \psi \right)
_{a}\bar{p}_{a}+\left( \partial _{0}\bar{\psi}\right)_{a}p_{a}- \mathcal{H_{C}}\Big] \Big) , \end{eqnarray} where $\mathcal{H_{C}}$ is given by Eq.(\ref{eq
1.16}) and $\mathcal{D}\mu $ is the Liouville measure given by \begin{eqnarray} \mathcal{D}\mu \left[ {\psi ,\bar{\psi},A_{\mu },\theta }\right] =\mathcal{D}
\theta \mathcal{D}\pi _{\theta }\mathcal{D}\pi ^{\mu }\mathcal{D}A_{\mu } \mathcal{D}\bar{\psi}\mathcal{D}\psi \mathcal{D}\bar{p}\mathcal{D}p~\delta \left(
\Sigma _{a}\right) \det\left\vert \left\{ \chi _{k},\varphi _{j}\right\} _{B} \right\vert~\det \left\vert \left\{ \phi _{a},\bar{\phi} _{b}\right\}
_{B}\right\vert^{-1/2} \end{eqnarray} being $\Sigma =\left\{ \varphi _{k},\chi _{k},\phi _{a},\bar{\phi} _{a}\right\} $ the complete set of constraints of
theory.

By performing the integrations over the canonical conjugate momenta and after some manipulations we obtain the following expression for the transition amplitude
\begin{eqnarray} Z &=&\int \mathcal{D}\theta \mathcal{D}A_{\mu }\mathcal{D}\bar{\psi}\mathcal{ D}\psi \det \left\vert -\left( \partial _{1}\right)
^{2}\right\vert \delta \left( \partial _{1}A_{1}\right)  \label{eq 2.1} \\ &&~\times \exp \left( i\int d^{2}x\left[ \bar{\psi}\left( i\partial
\!\!\!/+A\!\!\!/-m\right) \psi -\frac{1}{4e^{2}}F_{\mu \nu }F^{\mu \nu }+ \frac{1}{2g}\left( A_{\mu }-\partial _{\mu }\theta \right) ^{2}\right] \right) .
\notag \end{eqnarray} It is well known that if the covariance of a transition amplitude is explicit, the calculation process becomes more manageable, however,
the transition amplitude obtained above is not explicitly covariant. The procedure to pass from a noncovariant gauge to a covariant one can be performed using
the Faddeev-Popov-De Witt ansatz \cite{16a,16b}. In our case we choose the $R_{\xi }-$gauge \begin{equation} R_{\xi }\left[ A,\theta \right] =\partial _{\mu
}A^{\mu }+\frac{\xi }{g} \theta ,  \label{eq 2.2} \end{equation} which the main purpose is decouple the $\theta -$field from the other fields. Therefore,
$R_{\xi }-$gauge allows us factorize the $\theta -$ contribution and rewrite the vacuum-vacuum transition amplitude in the following way \begin{equation} Z=\int
\mathcal{D}\theta \mathcal{D}A_{\mu }\mathcal{D}\bar{\psi}\mathcal{D} \psi \det \left\vert \square +\frac{\xi }{g}\right\vert \exp \left( i\int d^{2}x\left[
\mathcal{L}_{\psi ,A}+\mathcal{L}_{\theta }\right] \right) , \label{eq 2.4} \end{equation} where the Lagrangian densities $\mathcal{L}_{\psi ,A}$ and
$\mathcal{L} _{\theta }$ are defined by \begin{equation} \mathcal{L}_{\psi ,A}= \bar{\psi}\left( i\partial \!\!\!\!/+A\!\!\!\!/-m\right) \psi
-\frac{1}{4e^{2}}F_{\mu \nu }F^{\mu \nu }+ \frac{1}{2g}A_{\mu }A^{\mu }-\frac{1}{2\xi }\left( \partial _{\mu }A^{\mu }\right) ^{2},  \label{eq 2.5}
\end{equation} and \begin{equation} \mathcal{L}_{\theta }= \frac{1}{2g}\left( \partial _{\mu }\theta \right) \left( \partial ^{\mu }\theta \right) -\frac{\xi
}{2g^{2}} \theta^{2}, \label{eq 2.6} \end{equation} we observe that in the $R_{\xi }-$gauge the scalar field gains a mass which is gauge-fixing parameter
dependent: $m_{\theta }^{2}= \displaystyle\frac{ \xi }{g^{2}}$. Also, we can see through the Eq.(\ref{eq 2.6}) that Faddeev-Popov-De Witt's ghosts are decoupled
from the gauge and fermion fields, then the determinant contribution could be absorbed into of a normalization constant.

\subsection{The generating functional}

Now we will determine the generating functional starting from which we will compute the Green's functions as gauge and fermions field propagators and vertex
functions and also we will get the generalized WT identities.

Starting from the well-defined transition amplitude in (\ref{eq 2.4} ) we define the generating functional by \begin{equation} \mathscr{Z}\left[ \eta
,\bar{\eta},J^{\mu },K\right] =N\int \mathcal{D} \theta \mathcal{D}A_{\mu }\mathcal{D}\bar{\psi}\mathcal{D}\psi \exp \Big( i\int d^{2}x\Big[\mathcal{L}_{\psi
,A}+\mathcal{L}_{\theta }+\bar{\eta}\psi + \bar{\psi}\eta +J_{\mu }A^{\mu }+K\theta \Big]\Big).\quad\label{eq 2.7} \end{equation} Where $N$ is a normalization
factor defined such that $\mathscr{Z}[0,0,0,0]=1 $, hence, from the generating functional (\ref{eq 2.7}) we can compute all the Greens's functions,
Schwinger-Dyson-Fradkin equations and Ward-Takahashi identities of the model. In the $R_{\xi }-$gauge, the full generating functional can be factorized in the
product of two generating functionals as shown to follow \begin{equation} \mathscr{Z}\left[ \eta ,\bar{\eta},J^{\mu },K\right] =\tilde{\mathscr{Z}} \left[
K\right] \times \bar{\mathscr{Z}}\left[ \eta ,\bar{\eta},J^{\mu } \right] .  \label{eq 2.8} \end{equation} where $\tilde{\mathscr{Z}}\left[ K\right] $ is a
free scalar generating functional which permits to see the $\theta -$ field decouples of the gauge and fermion fields. The interacting generating functional
$\bar{\mathscr{Z}} \left[ \eta ,\bar{\eta},J^{\mu }\right] $ is responsible to generate all the correlation functions between the $\psi ,\bar{\psi}$ and $A_{\mu
}$ fields. This separation will be essential in what follows.

First, we perform the integration of the fermionic fields which leads to the massive fermionic determinant, however, here we go to consider the massless case,
$m=0$, to avoid to lose the principal motivation in the study of two-dimensional models, to gain nonperturbative information about the system. Thus, in the
massless case, the gauge invariant fermionic determinant is given by \cite{17a,17b} \begin{equation} \det \left( i\partial \!\!\!/+A\!\!\!\!/\right) =\exp \left[
\frac{i}{2\pi } \int d^{2}z~A_{\mu }\left( z\right) \left( \eta ^{\mu \nu }-\frac{\partial _{z}^{\mu }\partial _{z}^{\nu }}{\square }\right) A_{\nu }\left(
z\right)  \right] .  \label{eq 2.9} \end{equation}

After the fermionic integration, we write the interacting generating functional $\bar{\mathscr{Z}}\left[ \eta ,\bar{\eta},J^{\mu } \right] $ only in terms of the
gauge field \begin{eqnarray} \bar{\mathscr{Z}}\left[ \eta ,\bar{\eta},J^{\mu }\right] = \int \mathcal{D} A_{\mu }\exp \Big( i\int d^{2}x\Big[ \frac{1}{2}A_{\mu
}B_{\xi }^{\mu \nu }A_{\nu }+J_{\mu }A^{\mu }-\int d^{2}y~\bar{\eta}\left( x\right) G\left( x,y;A\right) \eta \left( y\right) \Big] \Big) ,  \label{eq 2.10}
\end{eqnarray} where we have defined the differential operator $B_{\xi }^{\mu \nu }$ as \begin{equation} B_{\xi }^{\mu \nu }= \left( \frac{1}{\pi
}+\frac{1}{g}+\frac{\square }{e^{2}} \right) T^{\mu \nu }+\left( \frac{1}{g}+\frac{\square }{\xi }\right) L^{\mu \nu },  \label{eq2.11} \end{equation} with
projectors $T^{\mu \nu }$ and $L^{\mu \nu }$ \begin{equation} T^{\mu \nu }= \eta ^{\mu \nu }-\frac{\partial ^{\mu }\partial ^{\nu }}{ \square },\quad L^{\mu \nu
}= \frac{\partial ^{\mu }\partial ^{\nu }}{ \square }.  \label{eq 2.12} \end{equation}

The function $G\left( x,y;A\right) $ is the Green's function of Dirac equation, it is given by \begin{equation} G\left( x,y;A\right) =\exp \left[ -i\int
d^{2}z~s^{\mu }\left( z,x,y\right) A_{\mu }\left( z\right) \right] G_{F}\left( x-y\right) ,  \label{eq 2.13} \end{equation} with $s^{\mu }\left( z,x,y\right) $
\begin{equation} s^{\mu }\left( z,x,y\right) =\left( \partial _{z}^{\mu }+\gamma _{5}\tilde{ \partial}_{z}^{\mu }\right) \left[ D\left( z-x\right) -D\left(
z-y\right)  \right] . \end{equation} where $G_{F}\left( x-y\right) $ and $D\left( x-y\right) $ are the Dirac and Klein-Gordon-Fock free Green's functions,
respectively.

With these functionals in hands we are ready to compute the correlation functions.

\subsection{The Green's functions}

An important remark about the results of this section is their nonperturbative character. It is possible due to the exact evaluation of the massless fermionic
determinant (\ref{eq 2.9}) in (1+1)-dimensions guaranteeing the obtention of exact expressions for the Green's functions.

\subsubsection{The gauge field propagator}

The gauge field propagator is defined as \begin{equation} \mathscr{D}_{\mu \nu }^{\xi }\left( x-y\right) = \left\langle 0\left\vert T \left[ A_{\mu } \left(
x\right) A_{\nu }\left( y\right) \right]\right\vert 0\right\rangle =-\left. \frac{\delta ^{2}\bar{\mathscr{Z}} \left[ \eta ,\bar{ \eta},J_{\mu }\right] }{\delta
J^{\nu } \left( y\right) \delta J^{\mu }\left( x\right) }\right\vert _{\eta =\bar{\eta}=J_{\mu }=0}. \label{eq 2.14} \end{equation} In momentum space, the gauge
field propagator is expressed as \begin{equation} i\tilde{\mathscr{D}}_{\mu \nu }^{\xi }\left( k\right) = \displaystyle{\frac{
e^{2}}{k^{2}-\frac{e^{2}}{g}-\frac{e^{2}}{\pi }}}\eta _{\mu \nu }+f^{\xi }\left( k\right) k_{\mu }k_{\nu },  \label{eq 2.15} \end{equation} where we have
defined the function $f^{\xi }\left( k\right) $ by \begin{equation} f^{\xi }\left( k\right) = \frac{\xi }{k^{2}\left( k^{2}-\frac{\xi }{g} \right)
}-\frac{e^{2}}{k^{2}\left( k^{2}-\frac{e^{2}}{g}-\frac{e^{2}}{\pi } \right) }.  \label{eq 2.16} \end{equation}

From expression Eq.(\ref{eq 2.15}) to $\tilde{\mathscr{D}}_{\mu \nu }^{\xi }(k)$, we see clearly that the gauge field propagator does presents neither infrared
nor ultraviolet divergences and its high-energy behavior goes as $ k^{-2}$. Also, the transverse component of the gauge field propagator has a physical squared
mass given by $\displaystyle\frac{e^{2}}{g}+\frac{e^{2}}{ \pi }$, being the second contribution dynamically generated after fermionic fields quantization.

\subsubsection{The fermionic propagator}

The fermionic propagator is given by \begin{equation} \mathscr{S}^{\xi }\left( x-y\right) = \left\langle 0\left\vert T\left[ \psi \left( x\right)
\bar{\psi}\left( y\right) \right] \right\vert 0\right\rangle =-\left. \frac{\delta ^{2}\bar{\mathscr{Z}}\left[ \eta ,\bar{\eta},J_{\mu } \right] }{\delta \eta
\left( y\right) \delta \bar{\eta}\left( x\right) } \right\vert _{\eta =\bar{\eta}=J_{\mu }=0},  \label{eq 2.17} \end{equation}

Then, from the generating functional (\ref{eq 2.10}), we find the following expression to the full fermionic propagator \begin{equation} \mathscr{S}^{\xi }\left(
x-y\right) =i\exp \left( i\int \frac{d^{2}k}{\left( 2\pi \right) ^{2}}f^{\xi }\left( k\right) \left[ \frac{{}}{{}} \!1-e^{-ik\cdot \left( x-y\right) }\right]
\right) G_{F}\left( x-y\right) \label{eq 2.18} \end{equation} with $f^{\xi }\left( k\right) $ given by Eq.(\ref{eq 2.16}). Also, we evaluated the
Schwinger-Dyson-Fradkin equation to the fermionic propagator \begin{equation} i\partial \!\!\!/\mathscr{S}^{\xi }(x-y)+\gamma ^{\mu }\mathscr{G}_{\mu }^{\xi
}(x,y;z)=i\delta (x-y),  \label{eq 2.19} \end{equation} being $\mathscr{G}_{\mu }^{\xi }(x,y;z)$ given by Eq.(\ref{eq 2.22}). Due the exact evaluation of vertex
function $\mathscr{G}_{\mu }^{\xi }$, we rewrite the Schwinger-Dyson-Fradkin equation (\ref{eq 2.19}) in momentum space as \begin{equation}
\tilde{\mathscr{S}}^{\xi }(p)=\frac{i}{p\!\!\!/}-\frac{i}{p\!\!\!/}\int \frac{d^{2}k}{(2\pi )^{2}}k\!\!\!/f^{\xi }\left( k\right) \tilde{\mathscr{S}} ^{\xi
}(p-k),  \label{eq 2.20} \end{equation} starting from this expression we can construct a perturbative expansion for the fermionic propagator. It is easy to
verify that the power counting shows explicitly that it is free of ultraviolet divergences.

\subsubsection{The vertex function}

The complete vertex function defined by \begin{equation} \mathscr{G}_{\mu }^{\xi }\left( x,y;z\right) = \left\langle 0\left\vert T \left[ \psi \left( x\right)
\bar{\psi}\left( y\right) A_{\mu }\left( z\right) \right] \right\vert 0\right\rangle =i\left. \frac{\delta ^{3}\bar{ \mathscr{Z}}\left[ \eta ,\bar{\eta},J_{\mu
}\right] }{\delta J^{\mu }\left( z\right) \delta \eta \left( y\right) \delta \bar{\eta}\left( x\right) } \right\vert _{\eta =\bar{\eta}=J_{\mu }=0}
\end{equation} is obtained from the generating functional Eq.(\ref{eq 2.10}). After some calculation we obtain \begin{equation} \mathscr{G}_{\mu }^{\xi }\left(
x,y;z\right) =i\int \frac{d^{2}k}{\left( 2\pi \right) ^{2}}h_{\mu }(k)\left[ \frac{{}}{{}}\!e^{-ik\cdot \left( z-x\right) }-e^{-ik\cdot \left( z-y\right)
}\right] \mathscr{S}^{\xi }\left( x-y\right) ,  \label{eq 2.22} \end{equation} being $\mathscr{S}^{\xi }\left( x-y\right) $ the fermionic propagator given by
Eq.(\ref{eq 2.18}) and the function $h_{\mu }(k)$ is defined as \begin{equation} h_{\mu }(k)=-\frac{\xi }{k^{2}\left( k^{2}-\frac{\xi }{g}\right) }k_{\mu }-
\frac{e^{2}}{k^{2}\left( k^{2}-\frac{e^{2}}{g}-\frac{e^{2}}{\pi }\right) } \gamma _{5}\tilde{k}_{\mu }.  \label{eq 2.23} \end{equation}

To show that the vertex function (\ref{eq 2.22}) is free from ultraviolet divergences in the same way that fermionic propagator, it is convenient to write the
vertex function in Fourier space \begin{equation} \tilde{\mathscr{G}}_{\mu }^{\xi }\left( p,q;k\right) =i\left( 2\pi \right) ^{2}h_{\mu }(k)\left[
\tilde{\mathscr{S}}^{\xi }\left( p+k\right) -\tilde{ \mathscr{S}}^{\xi }\left( p\right) \right] \delta \left( p+k+q\right) , \label{eq 2.24} \end{equation}
where $\tilde{\mathscr{S}}^{\xi }\left( p\right) $ is the fermionic propagator (\ref{eq 2.18}) written in momentum space. We have seen that the fermionic
propagator is a finite function, hence, through Eq. (\ref{eq 2.24} ) we can conclude that the vertex function also will be finite.

\subsection{The Ward--Takahashi identities}

In this section we will derive the generalized WT identities satisfied by the $1PI$ functions. Such identities are a direct manifestation of the gauge symmetry
of GTM at quantum level. Firstly, we compute the general WT identity, in the $R_{\xi }-$gauge, satisfied by the 1PI generating functional Eq.(\ref{eq 2.27}). The
WT identity for the 1PI 2-point function of the gauge field shows the transverse character of the gauge field propagator. The next WT identity relates the 1PI
3-point vertex function with 1PI 2-point fermionic function, being it known as main WT identity. Also, we will show from the property Eq.(\ref{eq 2.8}) how the
effective action might be separated in two sectors, leading to two distinct equations. Thus, in order to get the identities of model, we firstly perform the
gauge transformations given in (\ref{eq 1.2}) into the generating functional Eq.( \ref{eq 2.7}). It is easy to note that the gauge-fixing and source terms in
the action (\ref{eq 2.7}) are not invariant under these local transformations, then, taking the strong hypothesis that the gauge transformation parameter
$\lambda \left( x\right) $ is an infinitesimal function, it allows us to obtain the WT identity satisfied by $\mathscr{Z} \left[ \eta
,\bar{\eta},J^{\mu},K\right] $ \begin{equation} \left[ \bar{\eta}\frac{\delta }{\delta \bar{\eta}\left( x\right) }-\eta \frac{\delta }{\delta \eta \left(
x\right) }+\frac{i}{g}\left( \square + \frac{\xi }{g}\right) \frac{\delta }{\delta K}+\frac{i}{\xi }\left( \square + \frac{\xi }{g}\right) \partial _{\mu
}\frac{\delta }{\delta J_{\mu }} -\partial _{\mu }J^{\mu }+K\right] \mathscr{Z}=0.  \label{eq 2.25} \end{equation}

Now, we introduce the generating functional of the connected Green's functions, $W\left[ \eta ,\bar{\eta},J^{\mu },K\right] $, defined as $ W=-i\ln
\mathscr{Z}$. Thus, the equation Eq.(\ref{eq 2.25}) allows to write the WT which is satisfied by the generating functional of the connected Green's functions
\begin{equation} i\bar{\eta}\frac{\delta W}{\delta \bar{\eta}}-i\eta \frac{\delta W}{\delta \eta }-\frac{1}{g}\left( \square +\frac{\xi }{g}\right) \frac{\delta
W}{ \delta K}-\frac{1}{\xi }\left( \square +\frac{\xi }{g}\right) \partial _{\mu }\frac{\delta W}{\delta J_{\mu }}-\partial _{\mu }J^{\mu }+K=0. \label{eq 2.26}
\end{equation}

Immediately, we define the generating functional of the 1PI functions, $ \Gamma \left[ \psi ,\bar{\psi},A_{\mu },\theta \right] $, by means of the following
Legendre transformation \begin{equation} \Gamma {\left[ \psi ,\bar{\psi},A_{\mu },\theta \right] }=W\left[ {\eta , \bar{\eta},J_{\mu },K}\right] -\int
d^{2}x\left( \frac{{}}{{}}\!\bar{\eta} \psi +\bar{\psi}\eta +J_{\mu }A^{\mu }+K\theta \right) ,  \label{eq 2.27} \end{equation} where following functional
relations are satisfied \begin{eqnarray} \frac{\delta W}{\delta \bar{\eta}(x)} &=&\psi (x)~,~~\frac{\delta W}{\delta \eta (x)}=-\bar{\psi}(x)~,~~\frac{\delta
W}{\delta K(x)}=\theta (x)~,~~\frac{ \delta W}{\delta J_{\mu }(x)}=A^{\mu }(x)~, \\ &&  \notag \\ \frac{\delta \Gamma }{\delta \psi (x)}
&=&\bar{\eta}(x)~,~\frac{\delta \Gamma }{\delta \bar{\psi}(x)}=-\eta (x)~,~\frac{\delta \Gamma }{\delta \theta (x)}=-K(x)~,~\frac{\delta \Gamma }{\delta A^{\mu
}(x)}=-J_{\mu }(x). \end{eqnarray}

By using the relations above, we obtain from Eq. (\ref{eq 2.26}) the general expression for the 1PI WT identities \begin{equation} i\frac{\delta \Gamma }{\delta
\psi }\psi -i\frac{\delta \Gamma }{\delta \bar{ \psi}}\bar{\psi}-\frac{1}{g}\left( \square +\frac{\xi }{g}\right) \theta - \frac{1}{\xi }\left( \square
+\frac{\xi }{g}\right) \partial ^{\mu }A_{\mu }+\partial _{\mu }\frac{\delta \Gamma }{\delta A_{\mu }}-\frac{\delta \Gamma }{\delta \theta }=0.  \label{eq 2.28}
\end{equation}

However, before continuing with the derivation of the identities of WT, we need to attend the following property: The equation (\ref{eq 2.8}) shows that the
functional generator can be written in the following way \begin{equation} \mathscr{Z}\left[ \eta ,\bar{\eta},J^{\mu },K\right] =\tilde{\mathscr{Z}} \left[
K\right] \times \bar{\mathscr{Z}}\left[ \eta ,\bar{\eta},J^{\mu } \right] ,  \notag \end{equation} and the immediate consequence is to write $W=-i\ln
\mathscr{Z}$, the full generating functional of the connected Green's functions as \begin{equation} W{\left[ \eta ,\bar{\eta},J_{\mu },K\right]
}=\tilde{W}{\left[ K\right] }+ \bar{W}{\left[ \eta ,\bar{\eta},J_{\mu }\right] },  \label{eq 2.29} \end{equation} and by putting it in (\ref{eq 2.27}) we find
\begin{equation} \Gamma {\left[ \psi ,\bar{\psi},A_{\mu },\theta \right] }=\tilde{\Gamma}{ \left[ \theta \right] }+\bar{\Gamma}{\left[ \psi ,\bar{\psi},A_{\mu
}\right] }.  \label{eq 2.30} \end{equation} Hence, we can write the Eq. (\ref{eq 2.28}) as two decoupled equations \begin{equation} i\frac{\delta
\bar{\Gamma}}{\delta \psi }\psi -i\frac{\delta \bar{\Gamma}}{ \delta \bar{\psi}}\bar{\psi}-\frac{1}{\xi }\left( \square +\frac{\xi }{g} \right) \partial ^{\mu
}A_{\mu }+\partial _{\mu }\frac{\delta \bar{\Gamma}}{ \delta A_{\mu }}=-C_{1},  \label{eq 2.31} \end{equation} and \begin{equation} \frac{1}{g}\left( \square
+\frac{\xi }{g}\right) \theta +\frac{\delta \tilde{ \Gamma}\left[ \theta \right] }{\delta \theta }=C_{1},  \label{eq 2.32} \end{equation} where $C_{1}$ is a
constant separation. It is easy to compute the solution of Eq.(\ref{eq 2.32}), and it is given by \begin{equation} \tilde{\Gamma}{\left[ \theta \right] }=\int
d^{2}x\left[ \frac{1}{2g}\left( \partial _{\mu }\theta \right) ^{2}-\frac{\xi}{2g^{2}} \theta ^{2} +C_{1}\theta \right] ,  \label{eq 2.33} \end{equation} which
is the action for scalar sector of theory. At this point, it is worthwhile to note that in the $R_\xi-$ gauge (\ref{eq 2.6}), we have $ \left\langle \theta
\right\rangle =0$ therefore $C_{1}=0$.

The first identity comes from applying the derivative with respect of $ A_{\nu }(y)$ on Eq.(\ref{eq 2.31}), this leads to the following equation for the 1PI\
2-point function of the gauge field, $\bar{\Gamma} _{\mu \nu }$, \begin{equation} \partial _{\mu }\bar{\Gamma}^{\mu \nu }\left( x-y\right) -\frac{1}{\xi }
\left( \square +\frac{\xi }{g}\right) \partial ^{\nu }\delta \left( x-y\right) =0,  \label{eq 2.34} \end{equation} which when it is written in momentum space
\begin{equation} k_{\mu }\bar{\Gamma}^{\mu \nu }\left( k\right) =\frac{1}{\xi }\left( \frac{ \xi }{g}-k^{2}\right) k^{\nu },  \label{eq 2.35} \end{equation}
shows clearly the transverse character of 1PI 2-point function of gauge field.

Now, the next WT identity relates the 1PI 3-point vertex function with a 1PI 2-point fermionic function, performing the $\psi (y)$ and $\bar{\psi}(z)$
derivatives on Eq.(\ref{eq 2.31}) we find the expression \begin{equation} i\bar{\Gamma}\left( x-z\right) \delta \left( z-x\right) +i\bar{\Gamma}\left( x-y\right)
\delta \left( x-y\right) +\partial _{\mu }\bar{\Gamma}^{\mu }\left( z,y;x\right) =0,  \label{eq 2.36} \end{equation} which in momentum space read as
\begin{equation} k^{\mu }\bar{\Gamma}_{\mu }^{\xi }\left( p;k\right) =\bar{\Gamma}^{\xi }\left( p\right) -\bar{\Gamma}^{\xi }\left( p+k\right) ,  \label{eq 2.37}
\end{equation} The functionals of r.h.s. are related with complete fermion propagator Eq.( \ref{eq 2.18}) by a functional identity
$\tilde{\mathscr{S}}(p)\bar{\Gamma} (p)=i$. It tells us that as the 1PI\ 2-point function of the fermion field is finite then the 1PI 3-point vertex function
will also be finite.

\section{Strong coupling isomorphisms of gauged Thirring model}

One of the goals of Kondo \cite{12a,12b,12c} was to analyze at classical level the behavior of the gauged Thirring model (in unitary gauge, $\theta =0$)  {in its
strong coupling limits recovering} both the Schwinger and the Thirring models, respectively. Now, we will go to establish in a nonperturbative way the strong
coupling limits in the quantum regimen. First, we will proceed to perform the analysis of the strong coupling limit $g\rightarrow \infty $ to get the quantum
Schwinger model. Next, it will perform the limit $e\rightarrow \infty $ to obtain the Thirring model which is a more complex issue.

\subsection{The isomorphism between GTM and SM}

The quantum isomorphism between the Green's functions of the gauged Thirring model and those of the Schwinger model are obtained in the limit $ g\rightarrow
\infty $. We show some achievements of this isomorphism between the fundamental Green's functions and the 1PI WT identities of the respective models:

\begin{enumerate} \item[(a)] By applying the limit in gauge field propagator (\ref{eq 2.15}), we obtain \begin{equation} i\tilde{\mathscr{D}}^{\mu \nu }\left(
k\right) =\frac{e^{2}}{k^{2}-m_{s}^{2}} T^{\mu \nu }\left( k\right) +\frac{\xi }{k^{2}}L^{\mu \nu }\left( k\right) , \label{eq 3.1} \end{equation} where
$m_{s}^{2}=\displaystyle{\frac{e^{2}}{\pi }}$ is the Schwinger mass.

\item[(b)] In fermionic propagator (\ref{eq 2.18}), the limit results in \begin{equation} \mathscr{S}^{\xi }(x-y)=i\exp \left\{ i\int \frac{d^{2}k}{(2\pi
    )^{2}}\frac{1 }{k^{2}}\left( \frac{\xi }{k^{2}}-\frac{e^{2}}{k^{2}-m_{s}^{2}}\right) \left[ \frac{{}}{{}}\!1-e^{-ik\cdot (x-y)}\right] \right\}
    G_{F}(x-y). \label{eq 3.2} \end{equation}

\item[(c)] For vertex function (\ref{eq 2.22}), the limit gives \begin{equation} \mathscr{G}_{\mu }^{\xi }(x,y;z)=\left(\frac{\xi }{\square _{z}}
    \partial_{\mu }^{z} +\frac{e^{2}}{\square _{z}+m_{s}^{2}}\gamma _{5} \tilde{ \partial}_{\mu }^{z}\right) \left[
    D\left(z-x\right)-D\left(z-y\right)\right] \mathscr{S}^{\xi }(x-y),  \label{eq 3.3} \end{equation} with $\mathscr{S}^{\xi }$ given by Eq.(\ref{eq 3.2}).
    \end{enumerate}

If we make the following redefinitions in the gauged Thirring model \begin{equation} A_{\mu }\rightarrow eA_{\mu },\quad \xi \rightarrow e^{2}\xi , \label{eq
3.4} \end{equation} the above equations reproduce exactly the Green's functions of the Schwinger model \cite{5}.

The general Ward-Takahashi identity (\ref{eq 2.31}), in the limit $ g\rightarrow \infty $ goes to \begin{equation} i\frac{\delta \bar{\Gamma}}{\delta \psi }\psi
-i\frac{\delta \bar{\Gamma}}{ \delta \bar{\psi}}\bar{\psi}-\frac{1}{e}\frac{\square }{\xi }\partial _{\mu }A^{\mu }+\frac{1}{e}\partial _{\mu }\frac{\delta
\bar{\Gamma}}{\delta A_{\mu }}=0,  \label{eq 3.5} \end{equation} which is exactly the WT for the Schwinger model.

Therefore, we can conclude that at quantum level the limit $g\rightarrow \infty $ is well-defined given a exact isomorphism between the gauged Thirring and
Schwinger models.

\subsection{The isomorphism between GTM and TM}

At first sight, naively, we can perform the limit $e\rightarrow \infty $ in all the GTM Green's functions previously computed and  {immediately to} eliminate the
dependence in the gauge-fixing parameter by taking the limit $ \xi \rightarrow \infty $ (unitary gauge). However, we must observe that in the TM does not exist
a gauge field therefore the GTM Green's function of operators containing only fermionic fields will have a correspondence with those of the TM already computed.
Thus, by taking the limits in the fermionic propagator (\ref{eq 2.18}) we obtain \begin{equation} \mathscr{S}\left( x-y\right) =i\exp \left(
-i\frac{g^{2}}{\left( \pi +g\right) }\int \frac{d^{2}k}{\left( 2\pi \right) ^{2}}\frac{1-e^{-ik\cdot \left( x-y\right) }}{k^{2}}\right) G_{F}\left( x-y\right)
\label{eq 3.6} \end{equation} which reproduces exactly the fermionic propagator of the Thirring model such as shown in \cite{18}.

Although we have calculated three correlation functions for GTM the equivalence with Thirring model, for now, is only between the pure fermionic Green's
functions, due to non-existence of a dynamical gauge field in TM. However, there is a vector field which is dynamical, the fermionic current. Thus, we can
establish the isomorphism between the Green's function involving fermionic fields and fermionic currents, as for example, the current propagator, $\langle
0|T[j_{\mu }(x)j_{\nu }(y)]|0\rangle $, and the vertex between the fermionic fields and fermionic current, $\langle 0|T[\psi (x) \bar{\psi}(y)j^{\mu
}(z)]|0\rangle $. To evaluate such correlation functions in GTM scenario it is necessary to introduce a source for the fermionic current $j^{\mu }$ into the
generating functional Eq.(\ref{eq 2.7} ), which we will denote by $C_{\mu }$, then, the generating functional is written now as $\mathscr{Z}\left[ \eta
,\bar{\eta} ,J_{\mu },C_{\mu },K \right] $. It is important to stress here that the addition of these source does not interfere in the results obtained until
now neither it harms the utility of the $R_{\xi }-$ gauge.

Now, we compute the relevant Green's functions of the gauged Thirring model containing only the fermionic current or/and fermionic.

Firstly, the GTM current propagator, $\mathscr{J}_{\mu \nu }\left( x-y\right) =\left\langle 0\left\vert T\left[ j_{\mu }\left( x\right) j_{\nu }\left( y\right)
\right] \right\vert 0\right\rangle $, expressed in momentum space is \begin{equation} i\mathscr{J}_{\mu \nu }\left( k\right) =\frac{1}{\pi
}\frac{k^{2}-\frac{e^{2} }{g}}{k^{2}-\frac{e^{2}}{\pi }-\frac{e^{2}}{g}}T_{\mu \nu },  \label{eq 3.8} \end{equation} which in the limit $e\rightarrow \infty $\
$\left( \xi \rightarrow \infty \right) $ reduces to \begin{equation} i\mathscr{J}_{\mu \nu }\left( k\right) =\frac{1}{\pi +g}T_{\mu \nu }. \label{eq 3.9}
\end{equation}

Next, the GTM 3-point Green's function, $\mathscr{H}^{\mu }\left( x,y,z\right) =\left\langle 0\left\vert T\left[ \psi \left( x\right) \bar{\psi }\left( y\right)
j^{\mu }\left( z\right) \right] \right\vert 0\right\rangle $ , has the following expression \begin{equation} \mathscr{H}^{\mu }\left( x,y,z\right) =-\left(
\partial _{z}^{\mu }+\frac{ \square _{z}+\frac{e^{2}}{g}}{\square _{z}+\frac{e^{2}}{\pi }+\frac{e^{2}}{g} }\gamma _{5}\tilde{\partial}_{z}^{\mu }\right) \left[
D\left( z-x\right) -D\left( z-y\right) \right] \mathscr{S}^{\xi }\left( x-y\right) \label{eq 3.11} \end{equation} where $\mathscr{S}^{\xi }\left( x-y\right) $
is the GTM fermion propagator given by (\ref{eq 2.18}). Then, taking the limit $e\rightarrow \infty ~\left( \xi \rightarrow \infty \right) $ we find
\begin{equation} \mathscr{H}^{\mu }\left( x,y,z\right) =-\left( \partial _{z}^{\mu }+\frac{ \pi }{\pi +g}\gamma _{5}\tilde{\partial}_{z}^{\mu }\right) \left[
D\left( z-x\right) -D\left( z-y\right) \right] \mathscr{S}\left( x-y\right) , \label{eq 3.12} \end{equation} being $\mathscr{S}\left( x-y\right) $ the TM
fermionic propagator given by Eq.(\ref{eq 3.6}). The expressions (\ref{eq 3.9}) and (\ref{eq 3.12}) agree with that found for the Thirring model in Ref.
\cite{18}.

Therefore, we conclude that there is, at quantum level, an isomorphism between the gauged Thirring and Thirring models. Such isomorphism is a mapping between
Green's functions of composite field operators only with fermionic fields and established via the limit $e\rightarrow \infty $ $ \left( \xi \rightarrow \infty
\right) $. It guarantees the quantum equivalence between the gauged Thirring and massless Thirring models.

\section{Remarks and conclusions}

We have quantized by using functional techniques the bidimensional gauged Thirring model finding closed expressions for the complete Green's functions and WT
identities. Next, we have established the quantum isomorphisms between the GTM and the Schwinger and massless Thirring models  {taking the respective strong
coupling limits}. In this way, we have extended the classical isomorphisms established by Kondo.

We have begun the study by establishing the Hamiltonian structure of the GTM following Dirac's procedure for constrained systems. Next, we construct a explicitly
noncovariant vacuum-vacuum transition amplitude via the Faddeev-Senjanovic's method and via the Faddeev-Popov-De Witt ansatz we obtain a covariant transition
amplitude in the $R_{\xi }-$gauge. This particular gauge simplifies the computation of Green's functions and WT identities due to it allows to decouple the
$\theta -$field from the gauge and fermion fields.

We must stand out our restriction to massless fermions case allowing to compute exactly the bidimensional fermionic determinant. Such as it is well-known the
integration of the massless fermionic fields generates a dynamical mass contribution $\frac{e^{2}}{\pi }$ (Schwinger mechanism)  {for the GTM gauge field} such
as it can be observed in the pole structure of the gauge propagator (\ref{eq 2.15}). This restriction is benefit because allows to obtain exact expressions for
the complete Green's functions as, for example, the propagators and vertex functions computed here. Also, it permits to write explicitly the
Schwinger-Dyson-Fradkin equation for the fermionic propagator on account of the vertex function is computed exactly.

In the last section, we have shown the quantum behavior of the gauged Thirring model in the  {strong coupling limits} by establishing the quantum isomorphisms
between the GTM and the Schwinger and massless Thirring models. The first quantum isomorphism is obtained in the $g\rightarrow \infty $ limit in which the GTM
reproduces exactly in a nonperturbative sense as all the correlation functions as the Ward-Takahashi identities of the Schwinger model. The second quantum
isomorphism is a mapping between GTM and TM Green's functions of composite field operators only with fermionic fields and established via the limits
$e\rightarrow \infty $ $\left( \xi \rightarrow \infty \right) $. In this way, both isomorphisms guarantee the quantum equivalence between the gauged Thirring and
Schwinger and massless Thirring models.

To finalize, we present some interesting comments on future developments. We can tell that finite temperature formulation of gauged Thirring model is a natural
continuation of the work developed here. Since, the thermodynamic properties of the Schwinger and Thirring models have been studied for long years, it is natural
the question about the existence of the finite temperature isomorphisms between the GTM and them. Also, the bosonization of gauged Thirring model is another
interesting point that deserves to be more elaborated and studied, because the bosonization provides a powerful tool to obtain nonperturbative information of
two-dimensional field theories. Once these results of GTM are obtained, they can give more information, and then improve our understanding about the quantum
structure of the isomorphism between GTM and the Schwinger and Thirring models. Progress in these directions will be reported elsewhere.

\subsection*{Acknowledgements}

RB thanks CNPq for full support, RC thanks to CNPq, CAPES and FAPEMA for partial support and BMP thanks CNPq and CAPES for partial support.

\end{document}